# Role of Strain on Electronic and Mechanical Response of Semiconducting Transition-Metal Dichalcogenide Monolayers: an ab-initio study

David M. Guzman and Alejandro Strachan

*School of Materials Engineering and Birck Nanotechnology Center,*

*Purdue University, West Lafayette, Indiana, USA*

We characterize the electronic structure and elasticity of monolayer transition-metal dichalcogenides $MX_2$ (M=Mo, W, Sn, Hf and X=S, Se, Te) with 2H and 1T structures using fully relativistic first principles calculations based on density functional theory. We focus on the role of strain on the band structure and band alignment across the series 2D materials. We find that strain has a significant effect on the band gap; a biaxial strain of 1% decreases the band gap in the 2H structures, by as a much 0.2 eV in $MoS_2$ and $WS_2$, while increasing it for the 1T materials. These results indicate that strain is a powerful avenue to modulate their properties; for example, strain enables the formation of, otherwise impossible, broken gap heterostructures within the 2H class. These calculations provide insight and quantitative information for the rational development of heterostructures based on these class of materials accounting for the effect of strain.

Since the successful exfoliation of one-atom-thick materials [1] interest in two-dimensional (2D) materials has grown hand in hand with our ability to control their synthesis [2], the characterization of their unique properties [3] and their initial use in devices [4]. Many of the unique physical properties of this new class of materials result from quantum confinement and the number of multi-layers can be used to tune their properties. Strain also has a strong effect on materials properties and whether engineered or unwanted it should be accounted for in applications.[5] The significant efforts in basic and applied science around graphene have been accompanied by growing interest in developing a menu of 2D materials that would offer the variety in electronic properties (e.g. bandgaps, band offsets and carrier mobilities) needed for device applications. For example, type-II heterostructures would be interesting for photovoltaics and type-III heterostructures for mid-wave infrared light sources and band-to-band tunnel field effect transistors. Progress in the synthesis and characterization of new 2D materials would benefit significant from theoretical guidance regarding the properties of known and postulated materials.

Transition-metal dichalcogenides ($MX_2$) layered materials are an attractive starting point for 2D material design due to their wide variety of physical properties, ranging from semiconductors as in the case of $(Mo, W)X_2$ to superconductors like $NbS_2$. Some of these materials have been successfully produced as single layers[1,6,7]. Furthermore, these 2D materials are being integrated into devices; for example $MoS_2$ monolayer based transistors[8] and integrated circuits to perform logic operations[9]. Other applications for photovoltaic devices[10], vapor sensing[11], spontaneous water photo-spliting[7,12], and spintronics[13] have also been proposed and studied. Furthermore, it has been demonstrated that high yield and reproducibility of $MX_2$ single-layer nanosheets can be achieved from the bulk material through lithium intercalation in an electrochemical set up [14] which is attractive for their application in the electronic industry. The design of such devices and synthesis efforts would benefit enormously from an a priori knowledge of the key properties of possible 2D materials. Recent reports used density functional theory (DFT) to predict band alignment for a wide range of dichalcogenides [15] and [16]. In this Letter we use DFT to characterize the thermodynamic stability, elastic properties and role of strain on the band alignment of dichalcogenides. The results confirm the possibility of creating type-II and type-III heterostructures but place important constraints on the levels of strain admissible to achieve them.

Bulk Mo and W dichalcogenides crystallize in a 4H structure with space group $P6_3/mmc$ (194) that consists of X-M-X slabs weakly bonded though van der Waals interactions. The space group symmetry is reduced to P-6m2 (187) in the single-layer system due to the loss of inversion symmetry. Bulk Sn and Hf dichalcogenides adopt a 1T structure with space group P-3m1 (164) identical to cadmium iodide ($CdI_2$); again, the individual layers are held together through weakly by van der Walls forces. Regardless of the dimensionality the $MX_2$ layers consist of metal atoms sandwiched between chalcogens through ionic-covalent bonding forming a trigonal prismatic coordination for M=Mo and W, as shown in

Figure 1(a), and an octahedral coordination for M=Sn and Hf, as shown in Figure 1(b).

The electronic structure of single- and few- layer Mo and W dichalcogenides has been extensively studied[17,18,19] and recent studies have studied the effects of mechanical strain on the electronic structure [20,21] and the band offsets [12,22] of some $MX_2$ single layer systems. Transition-metal dichalcogenide monolayers with a 1T structure have been identified as promising materials for electronic and photochemical applications such as broken-gap tunnel field effect transistor [15] and water photo-splitting devices [16]. A significant challenge for such applications is that the use $MX_2$ monolayers in devices or their hetero-integration will almost invariably lead to mechanical strain due to the lattice parameter or thermal expansion coefficient mismatch. Consequently, the role of strain on band alignment and the stiffness of these materials are critical pieces of information for rational device design. In this Letter we characterize the band structure of $MX_2$ monolayers consisting of M=Mo, W, Sn, Hf and X=S, Se, and Te using DFT; we focus on the effect of biaxial strain on the band offsets.

*Simulation details.* All calculations are carried out using the all-electron, full-potential, linear augmented plane wave (FP-LAPW) method as implemented in the WIEN2k code[23]. The exchange-correlation potential was calculated using the generalized gradient approximation of Perdew, Burke, and Ernzerhof (PBE)[24]. The muffin-tin radii for the chalcogen atoms was taken as RMT =2.1 a.u., while for the Mo and W atoms RMT=2.3 a.u., and for the Sn and Hf atoms RMT=2.5 a.u. We set the parameter $RMTK_{max}$=7, where $K_{max}$ is a cutoff wave vector. The valence electrons wave functions inside the muffin-tin spheres are expanded in terms of spherical harmonics up to $l_{max}=10$, and in terms of plane waves with a wave vector cutoff $K_{max}$ in the interstitial region.

The two-dimensional structures are modeled as periodic slabs with a sufficiently large *c*-lattice constant (25 Å) to avoid interactions between adjacent layers. The in-plane lattice constant *a* and the internal position parameter *z* are optimized with a strict force convergence of 1 mRy/Bohr. The electronic integration is taken over a commensurate k-mesh of 105 x 105 in the two-dimensional Brillouin zone (figure 1(c) bottom) and the convergence of self-consistent field calculations is attained with a tolerance in total energy of 0.01mRy. Spin-orbit interaction is included in the Hamiltonian through the second variational method and is taken into account in the band structure. The vacuum level is taken as the zero energy for the band alignment calculation. We note that throughout this paper the band structures reported are obtained from the Kohn-Sham eigenvalues in the DFT calculations; while this is standard it is known to underestimate the band gap in semiconductors. Accurate GW calculations by *Liang et. al.* [22] in $MX_2$ monolayers predict an increase of the band gap of approximately 50% with respect to the standard DFT-GGA calculation. Furthermore, GW simulations confirm that the VBM and CBM follow the band-gap-center approximation, meaning that both band edges shift symmetrically in opposite directions with respect to the DFT-GGA calculated band-center. In discussing the implications of our work we will make use of this empirical correction.

*Structures and energetics.* As we reduce the dimensionality from bulk to single layer MX$_2$, the absence of a van der Waals interlayer interaction is expected to cause structural relaxations [25]. The equilibrium lattice parameters, band gap, cohesive energy, stiffness and Poisson's ratio of the various 2D materials are summarized in Table 1. Consistent with prior studies [12,20], we observe an expansion of about 1-2% in the in-plane lattice constant *a* when going from bulk to single layer. The separation between the metal and chalcogen layers remains essentially unchanged. We calculated the internal position parameter *z* in bulk compounds to be ~0.121 for M=Mo, W and ~0.258 for M=Sn, these are in good agreement with experimental values of 0.129 and 0.240, respectively[26]. Some of these materials have yet to be produced as single layers and in order to quantify whether thermodynamic factors could affect their fabrication we computed the relative stability with respect to the bulk (total energy difference per formula unit). The energy difference between single layer and bulk materials range between 0.19 and 0.25 eV per formula unit, see Table 1, indicating that the monolayer stability across of these materials is similar.

Table I. *Basic properties of MX$_2$ monolayers. $a_0$ refers to the optimized lattice parameter, $E_{gap}$ is the band gap calculated with the PBE exchange-correlation functional, d and i indicates direct or indirect band gap, respectively. $E_f$ is the formation energy per unit formula of the MX$_2$ single-layer with respect to the bulk material. I refers to the in-plane stiffness and $\nu$ the Poisson's ratio.*

| 2H Structures | $a_0$(Å) | $E_{gap}^{pbe}$ (eV) | $E_f$ (eV/u.f.) | $I(N/m)$ | $\nu$ |
|---|---|---|---|---|---|
| MoS$_2$ | 3.19 | 1.59(d) | 0.18 | 120.09 | 0.25 |
| MoSe$_2$ | 3.33 | 1.33(d) | 0.18 | 101.48 | 0.23 |
| MoTe$_2$ | 3.55 | 0.93(d) | 0.22 | 72.5 | 0.25 |
| WS$_2$ | 3.19 | 1.55(d) | 0.18 | 135.18 | 0.22 |
| WSe$_2$ | 3.32 | 1.27(d) | 0.19 | 112.35 | 0.20 |
| WTe$_2$ | 3.55 | 0.80(d) | 0.25 | 90.95 | 0.18 |
| **1T Structures** | $a_0$(Å) | $E_{gap}^{pbe}$ (eV) | $E_f$ (eV/u.f.) | $I(N/m)$ | $\nu$ |
| HfS$_2$ | 3.66 | 1.26(i) | 0.19 | 74.28 | 0.19 |
| HfSe$_2$ | 3.79 | 0.49(i) | 0.19 | 64.09 | 0.20 |
| SnS$_2$ | 3.71 | 1.55(i) | 0.19 | 66.04 | 0.24 |
| SnSe$_2$ | 3.87 | 0.73(i) | 0.20 | 55.18 | 0.24 |

*Band structure of monolayer TM dichalcogenides.* The absence of adjacent MX$_2$ layers induces strong modifications to the electronic structure of the monolayer systems due to quantum confinement. For systems consisting of M=Mo, W and X=S, Se, Te we observed a transition from indirect band gap in the bulk material to direct band gap in the single layer with valence band maxima (VBM) and conduction band minima (CBM) centered at the K point of the first Brillouin zone; Figures S2 and S3 in the supplementary material [27] show the electronic band structure of the MX$_2$ single-layer systems investigated in this work. This observation is consistent with experimental evidence showing the direct band gap character in MoS$_2$ monolayers[19,28] and prior calculations [12,18,20]. Although there is no experimental observation of direct band gap in WS$_2$ or WSe$_2$, we expect our results to be accurate on the basis of similarities of chemical composition and crystal structure of M=Mo and M=W systems. On the other hand, our simulations show that monolayers consisting of Sn and Hf atoms do not exhibit a direct band gap, as shown for $HfS_2$ and $SnS_2$, see Figure S3 in supplementary material. However, the VBM shifts from ~0.5(G->K) in bulk to ~0.3(G->M) in the single layer, while the CBM shifts from L to M.

*Band Alignment and the role of strain.* Now we turn attention to the band alignment of $MX_2$ monolayer systems and specifically the role of strain. Band alignment of all unstrained materials is presented in Figure S4 of the supplementary material.[27] Consistent with recent studies [12,15,16], we observe that the band edges of $MX_2$ monolayers with 2H structure increase as we move down in the chalcogen row, from S to Te. Overall, the VBM and CBM of $WX_2$ systems is energetically higher than $MoX_2$ structures[12]. For the $SnX_2$ monolayers the VBM increases in energy when X is changed from S to Se, however, the CBM remains essentially unchanged at about -5.1eV (remember these energies are referenced to the vacuum potential). In the case of $HfX_2$ single layer system the energy position of the VBM and CBM is consistent with the trends observed for $(Mo,W)X_2$ systems. As was recently reported [15], $(Hf, Sn)(S, Se)_2$ monolayers are good candidates to form type-III heterostructures with $Mo(Se, Te)_2$ and $W(Se, Te)_2$. The main challenge is now determining how strain affects the band alignments of the possible heterostructures.

To study to role of mechanical deformation we apply a uniform biaxial strain in the range of -5% to 5% to all systems. As shown in Figure 1(b), within the applied strain range all the monolayers studied, except for $HfSe_2$, remain semiconductors. Figure 1(b) shows that the electronic properties of these materials have a strong and complex dependence on strain. A 1% increase in the lattice parameter of Mo(W)S$_2$ monolayers results in a reduction of the bandgap of approximately 0.2 eV while for HfS$_2$ and HfSe$_2$ the same strain *increases* the bandgap by 0.1 eV. The bandgap of Sn(S,Se)$_2$ is significantly less sensitive to deformation: approximately 0.02 eV change per 1% strain. Tensile strain reduces the band gap of the 2H materials (arrows in Figure 1(b) indicate the relaxed lattice parameter) and this trend continues in compression up to approximately 1% (with the exception of WS$_2$). Interestingly, in compression this initial increase in band gap is followed by a decrease. This transition is associated with the shift from direct to indirect band gap, as shown in Figure 2 (top panels) for the selected case of MoTe$_2$. Consistent with this trend, in the indirect band gap 1T MX$_2$ monolayers tensile strain increases the band gap while compression diminishes it. This effect changes in SnS$_2$ when strained beyond approximately 3% (in tension); as shown in Fig. 2 (lower panels) a shift of the VBM from ~0.3($\Gamma$-M) to ~0.6($\Gamma$-K) is responsible for the band gap decreasing with additional tensile strain. Despite their similarities in crystal and electronic structure, the 1T 2D systems exhibit very different sensitivities to strain. Hf dichalcogenides have approximately 5x more strain sensitivity than the Sn systems. This directly correlates to the absolute position of the VBM and CBM as function of strain, which will be discuss next.

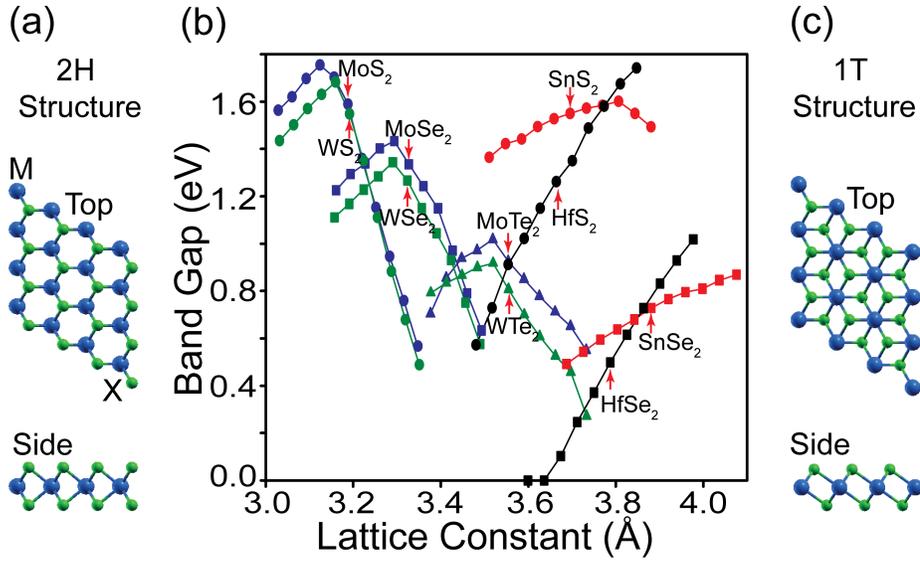

Figure 1. (Color Online) (a) Top and side view of 2H $MX_2$ monolayers (M=Mo, W and X=S, Se) crystal structure with space group symmetry P-6m2. (b) Band gap as function of the lattice constant for the different $MX_2$ monolayers studied. Each monolayer was strained biaxially in the range of +/-5%. The red arrow indicates the equilibrium lattice constant. (c) Top and side view of 1T $MX_2$ monolayers (M=Sn, Hf and X=S, Se) crystal structure with space group symmetry P-3m1.

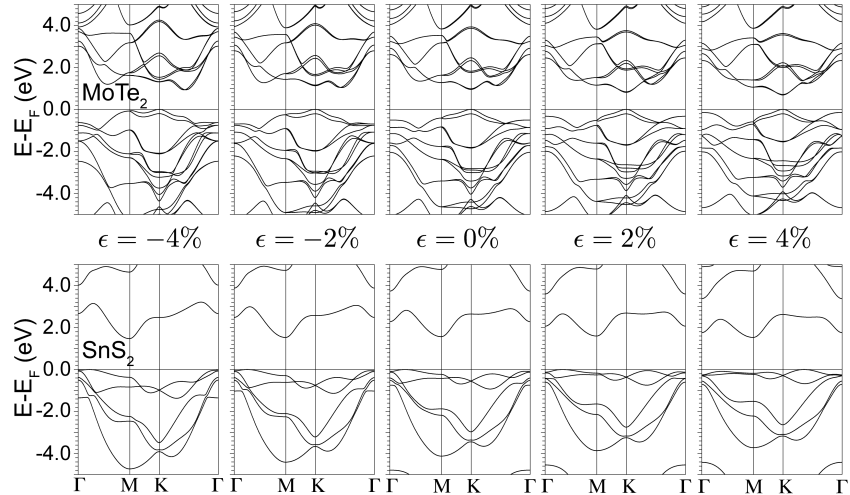

Figure 2. Band structure with SO interaction for selected cases showing the transition from direct-to-indirect band gap when compressive strain is applied. For the case of tensile strain the direct-to-indirect transition is more robust

In order to study band alignment in possible heterostructures, Figure 3 plots band edges energies with respect to vacuum for selected $MX_2$ monolayers as function of the lattice parameter. The significant sensitivity of the VBM and CBM energies to slight variations in the lattice parameter suggests significant flexibility in tuning band edge positions for specific applications but also point to the fact the small levels of strain can interfere with device performance. The 2H sulfide monolayers exhibit a minimum in the VBM at their equilibrium lattice parameter (arrows in Fig. 3); this minima move towards larger strains (in tension) as we move down the chalcogen row to Se and Te. A maximum in the CBM is observed in all 2H monolayers under approximately 2% compression; this maxima mark the transition from direct to indirect character of the band structure. On the other hand, the VBM and CBM of the 1T monolayer materials exhibit a monotonic behavior with respect to strain. The difference in behavior of the CBM with strain for $HfS_2$ and $SnS_2$ are responsible for the significant difference in strain sensitivity of these two materials.

*Discussion: heterostructure design*. Recent developments[29] in the fabrication of hybrid $MoS_2$-graphene[30] and $WS_2$-graphene[31] vertical heterostructures have driven much interest in the possibility of synthesizing free standing $MX_2$ heterostructures tailored to a wide variety of applications. In this context we discuss the possibility of forming type-III heterostructures with the studied $MX_2$ monolayers. In order to account for the known underestimation of the band gap in the Kohn Sham DFT-GGA band structures we include the GW-based 50% correction, discussed above. [22]

The strong dependence of band edges on strain provides additional flexibility in the design of heterostructures if strain could be independently controlled on the two materials but restricts options in epitaxial heterostructures, i.e. when the two materials share the same lattice parameter. While broken gap heterostructures cannot be formed with unstrained 2H materials, our results show that strain may enable them. For example, our calculations predict that $MoS_2$ or $WS_2$ in tension combined with $MoTe_2$ or $WTe_2$ under compression would form broken gap heterostructures even after the 50% GW correction is applied. This would occur for a lattice parameter of approximately 3.35Å and would require a significant strain on both materials (about 5%); while large this might be possible at the nanoscale and for 2D materials. [32] Broken gap heterostructures could also be formed between Mo and W tellurides and 1T materials studies over a larger range of lattice parameters and involving smaller strains.

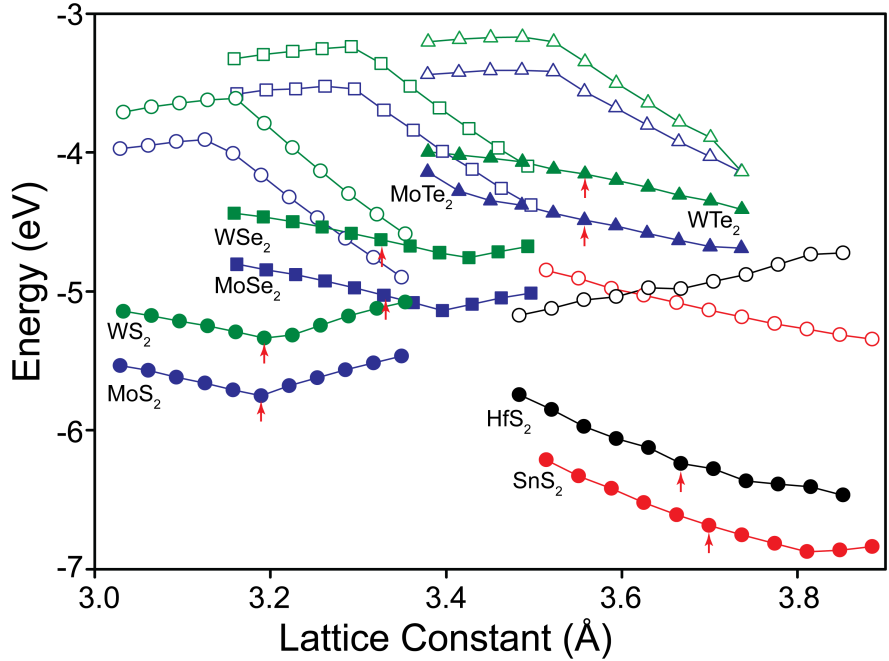

*Figure 3. (Color online) Absolute VBM and CBM position for selected cases as function of lattice constant. The vacuum level has been taken as reference and biaxial strain is varied from -5% to 5% in steps of 1% with respect to the equilibrium lattice constant for each system. The red arrows indicate the equilibrium lattice parameter for each material.*

*Stiffness of the 2D materials.* Finally, knowing the stiffness of the materials is critical to understand and engineer strain, yet experimental measurements for free standing 2D materials remain challenging.[33] We computed the 2D elastic stiffness of the all the monolayers studied from energy-strain relationships in the strain range -2% to 2% as described by *Topsakal et. al* [34]. The resulting values are plotted against lattice parameter in Figure 4. As expected, materials with smaller lattice parameters tend to be stiffer. We observe that within the 2H monolayers, the $WX_2$ systems are consistently stiffer than the $MoX_2$ monolayers. In both cases, as the chalcogen atom is changed from S to Te, the metal-chalcogen bond becomes longer and more ionic leading to a reduction in stiffness. Interestingly, the stiffness of 1T $MX_2$ single-layer systems follows the same trends as the 2H systems. HfX$_2$ monolayers exhibit a slightly higher stiffness than the SnX2 systems, but the overall stiffness of these systems is low ranging from 55 to 75 N/m.

In summary, we performed a systematic study of the electronic properties of monolayter transition metal dicalchogenides as a function of strain. While some of these 2D materials have not been experimentally realized yet their stability indicates no thermodynamic obstacles for their synthesis. The band alignment across the series shows significant flexibility in building heterostructures consisting of single-layers semiconducting materials of interest in a variety of applications. Our predictions indicate several options to create type-III, broken gap, heterostructures and that strain should be carefully considered in such designs.

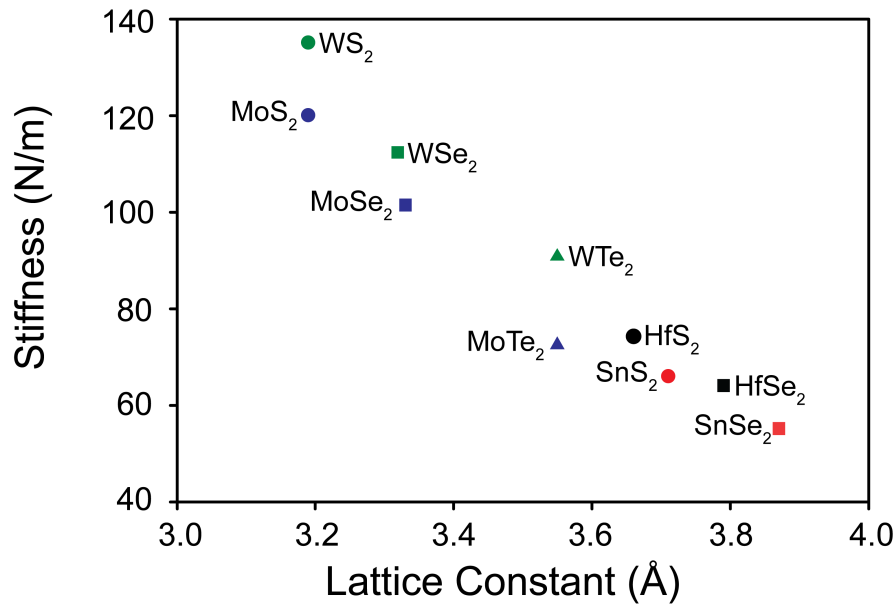

*Figure 4. (Color online) Calculated 2D stiffness as function of lattice parameter.*

# Supplementary Material for
# Role of Strain on Electronic and Mechanical Response of Semiconducting Transition-Metal Dichalcogenide Monolayers: an ab-initio study

David M. Guzman and Alejandro Strachan

School of Materials Engineering and Birck Nanotechnology Center,

Purdue University, West Lafayette, Indiana, USA

**Structure optimization**

All MX$_2$ monolayers were built based on the DFT relaxed geometries of their bulk systems. Table S1 shows a summary of the relaxed lattice parameters and band gaps of the bulk materials. For comparison, the experimental values have been included

*Table S1. (Color Online) Calculated and experimental lattice parameters and band bag energy of bulk transition-metal dichalcogenides. In all cases the band gap appears to be indirect.*

|  | PBE+vdWDF | | | Experimental | | |
|---|---|---|---|---|---|---|
| 2H Structures | $a(\text{Å})$ | $c(\text{Å})$ | $E_{gap}$ | $a(\text{Å})$ | $c(\text{Å})$ | $E_{gap}$ |
| MoS$_2$ | 3.21 | 12.44 | 0.85 | $3.16^a$ | $12.29^a$ | $1.23^a$ |
| MoSe$_2$ | 3.33 | 13.05 | 0.85 | $3.30^a$ | $12.94^a$ | $1.09^a$ |
| MoTe$_2$ | 3.54 | 14.01 | 0.68 | $3.52^a$ | $13.97^a$ | $1.00^a$ |
| WS$_2$ | 3.19 | 13.00 | 1.24 | $3.15^b$ | $12.32^b$ | $1.35^b$ |
| WSe$_2$ | 3.34 | 13.40 | 1.08 | $3.28^b$ | $12.96^b$ | $1.20^b$ |
| WTe$_2$ | 3.56 | 14.49 | 0.80 | – | – | – |
|  | PBE+vdWDF | | | Experimental | | |
| 1T Structures | $a(\text{Å})$ | $c(\text{Å})$ | $E_{gap}$ | $a(\text{Å})$ | $c(\text{Å})$ | $E_{gap}$ |
| SnS$_2$ | 3.63 | 5.9 | 1.52 | $3.64^c$ | $5.67^c$ | $2.34^d$ |
| SnSe$_2$ | 3.80 | 6.2 | 0.47 | $3.81^c$ | $6.14^c$ | $1.00^d$ |
| HfS$_2$ | 3.67 | 6.05 | 1.57 | $3.63^e$ | $5.88^e$ | $1.96^f$ |
| HfSe$_2$ | 3.78 | 6.64 | 0.89 | $3.74^e$ | $6.14^e$ | $1.14^g$ |

(a) Ref [1]
(b) Ref [2]
(c) Ref [3]
(d) Ref [4]
(e) Ref [5]
(f) Ref [5,6]
(g) Ref [5,7]

**Formation energy**

The formation energy of MX$_2$ single-layer systems, see Figure S1, was calculated with respect to the bulk material as

$$E_f = E_{monolayer} - E_{bulk}$$

Where $E_{monolayer}$ and $E_{bulk}$ are the total energies per formula unit of the monolayer and bulk systems, respectively. For the bulk systems we used a van der Waals functional to describe the nonlocal correlation energy as function of the electron density, as described by Dion et. al.[8] and implemented in the QUANTUM ESPRESSO code [9]. For completeness, we calculated the formation energy of graphene (C6) and hexagonal boron nitride (h-BN), which agree with previous calculations [10].

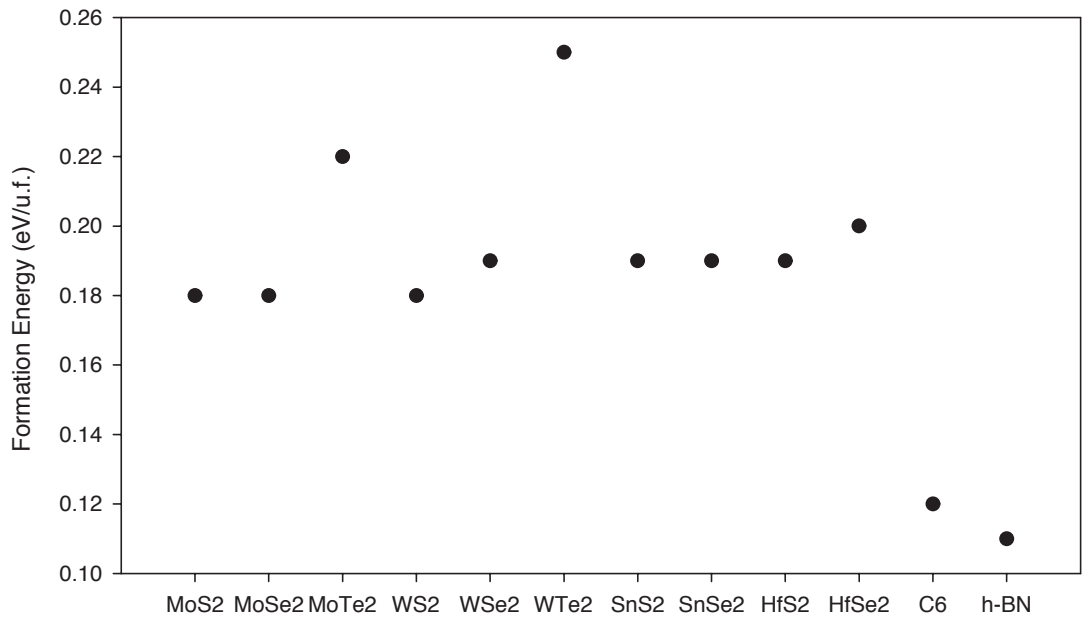

*Figure S1. Calculated formation energy of the studied transition-metal chalcogenide monolayers. For comparison we include the formation energies of graphene (C6) and hexagonal boron nitride (h-BN)*

**Electronic structure of MX$_2$ monolayers**

Figure S2 shows the electronic band dispersion and density of states of 2H transition-metal dichalcogenide monolayers calculated including the spin-orbit interaction. In all cases we observe a direct band gap centered at the K point in the two-dimensional Brillouin zone. On the other hand, the spin splitting of the valence band increases as the chalcogen atom is changed from S to Te, where systems with M=W exhibit a higher spin splitting as compared to monolayers consisting of M=Mo. Monolayers of WTe$_2$ exhibit the highest spin splitting ~400meV. Degeneracy of the conduction band close to its minima is not lifted due to spin-orbit induced spin splitting. From the partial density of states it is observed that the valence band maxima and conduction band minima are originated from the transition-metal atom states.

MX$_2$ monolayers with 1T structure do not exhibit direct band gap as shown in Figure S3. In the case of monolayers consisting of M=Sn, we observe the valence band maxima located at ~0.3G and the conduction band minima centered at the M symmetry point. Similarly, cases with M=Hf have the valence band maxima center at G and the conduction band minima at M. For both, M=Sn and Hf, the valence and conduction bands are originated from the metal atom, as seen from the partial density of states. Contrary to the 2H monolayers, the 1T monolayers do not exhibit a sizable spin splitting.

The origin of the observed spin-orbit induced spin splitting in 2H MX$_2$ monolayers has been attributed to a loss of inversion symmetry when the dimensionality is reduced from bulk 2H MX$_2$ to the monolayer [11].

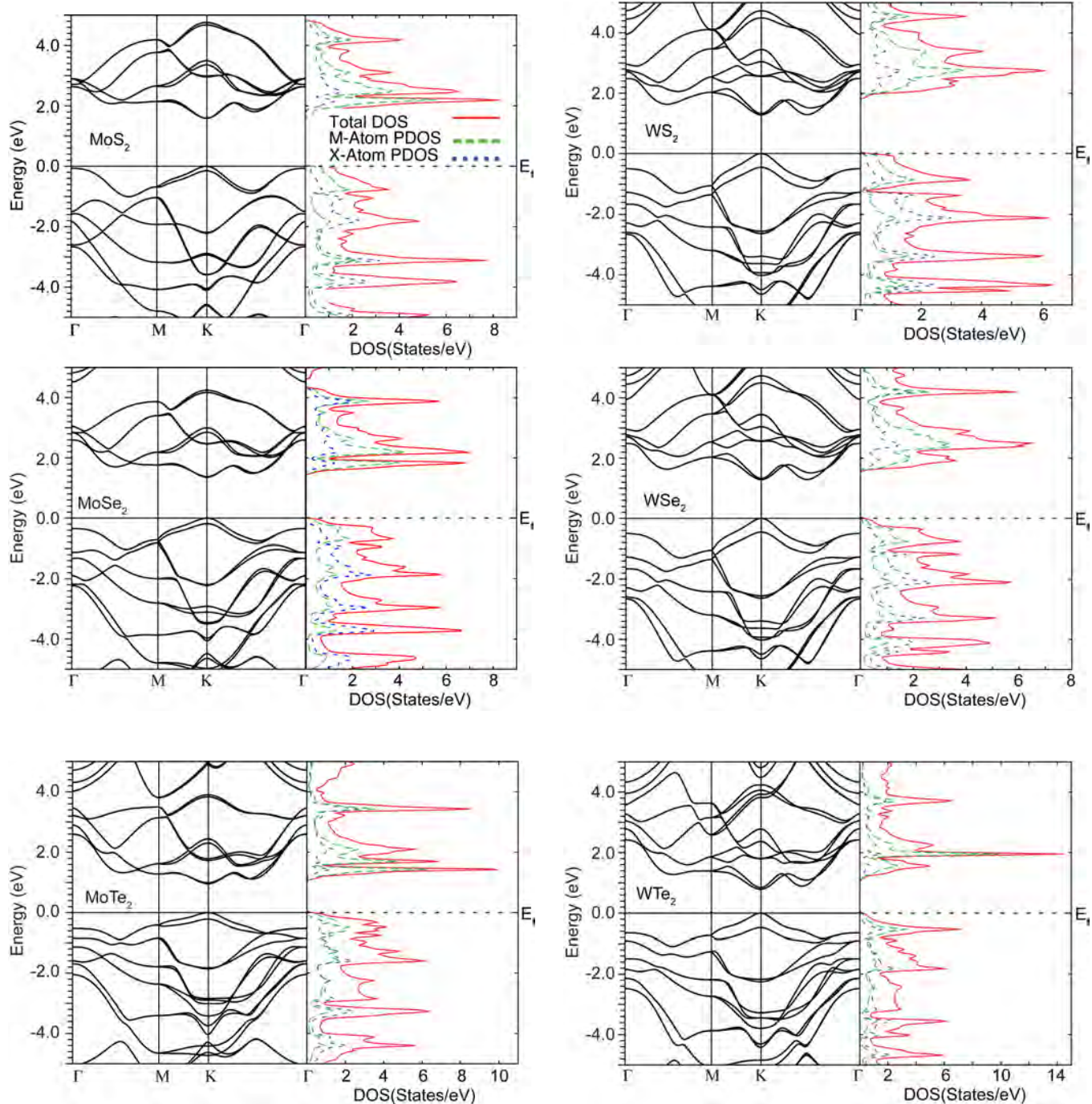

Figure S2. (Color Online) Calculated electronic band structure and partial density of states of transition-metal dichalcogenide monolayers with trigonal prismatic coordination. The spin-orbit interaction has been taken into account in the calculation. The Fermi energy is at 0 eV.

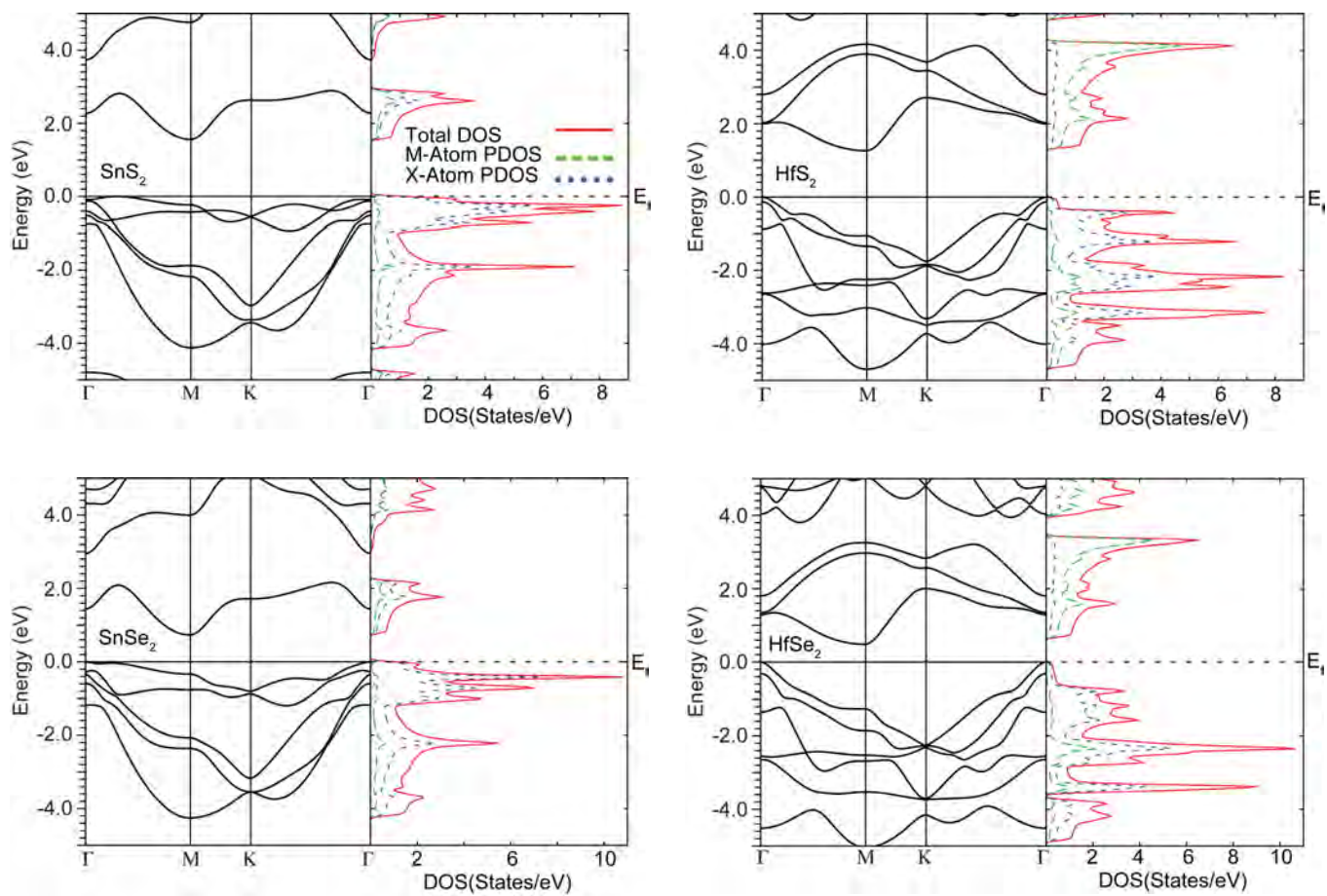

*Figure S3. (Color Online) Calculated electronic band structure and partial density of states of transition-metal dichalcogenide monolayers with octahedral coordination. The spin-orbit interaction has been taken into account in the calculation. The Fermi energy is at 0 eV*

**Unstrained band alignment of MX$_2$ monolayers**

For completeness we also carried out the band alignment calculation without spin-orbit interaction. We observe that the spin-orbit interaction affects the energetics of the VBM and CBM; specifically, the absolute position of the VBM is consistently higher when the SO interaction is turned on in the calculation, while the CBM remains basically unchanged. Moreover, the spin-orbit interaction is stronger in systems lacking inversion symmetry (P-6m2) than in centrosymmetric (P-3m1) structures[11]. The analysis presented in this work is based solely on the calculations with spin-orbit interaction.

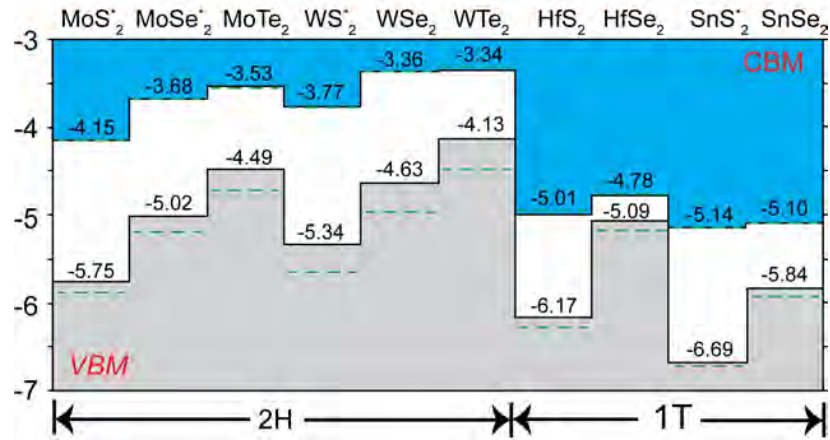

Figure S4. (Color Online) Band Alignment of MX$_2$ single-layer systems categorized by crystal structure. The black solid line represents the calculated band alignment with spin-orbit interaction, and the green dashed line represents the band alignment without spin-orbit interaction. The vacuum level was taken as 0eV. Those compounds with asterisk (*) have been produced experimentally as monolayers.[12]

## Role of strain on the band alignment of MX₂ monolayers

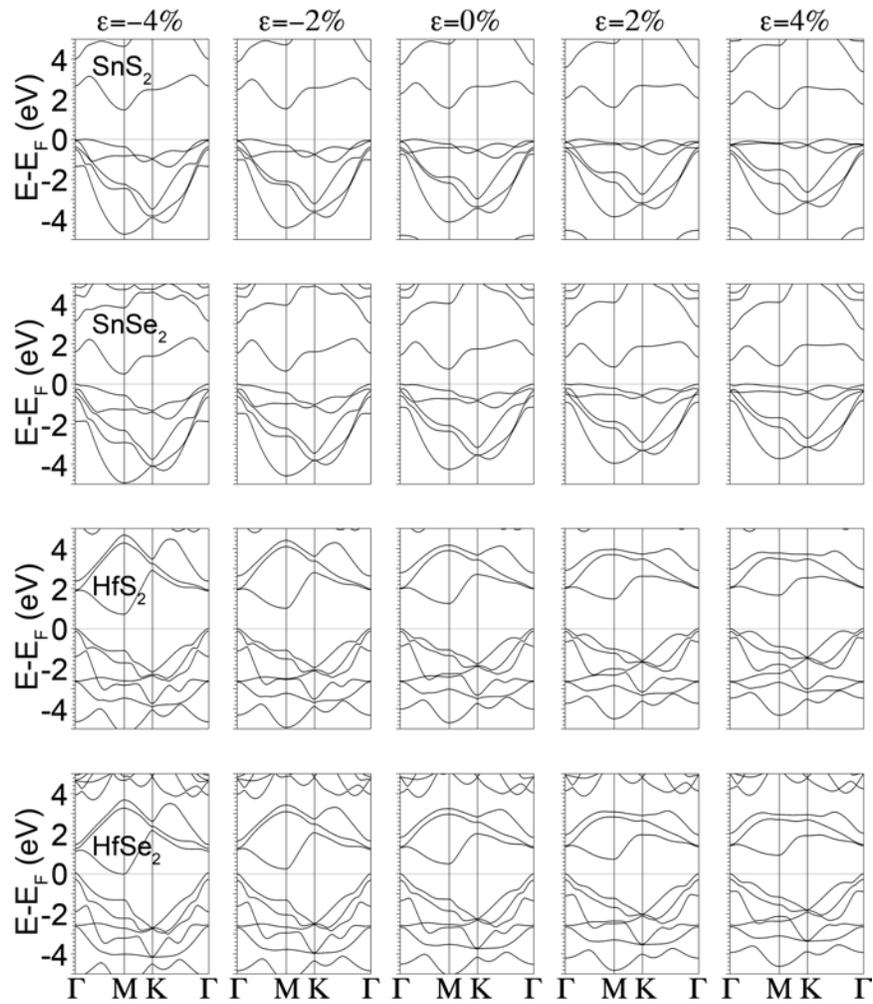

Figure S5. Band structure with SO interaction for transition-metal dichalcogenide single-layer systems with an octahedral coordination.

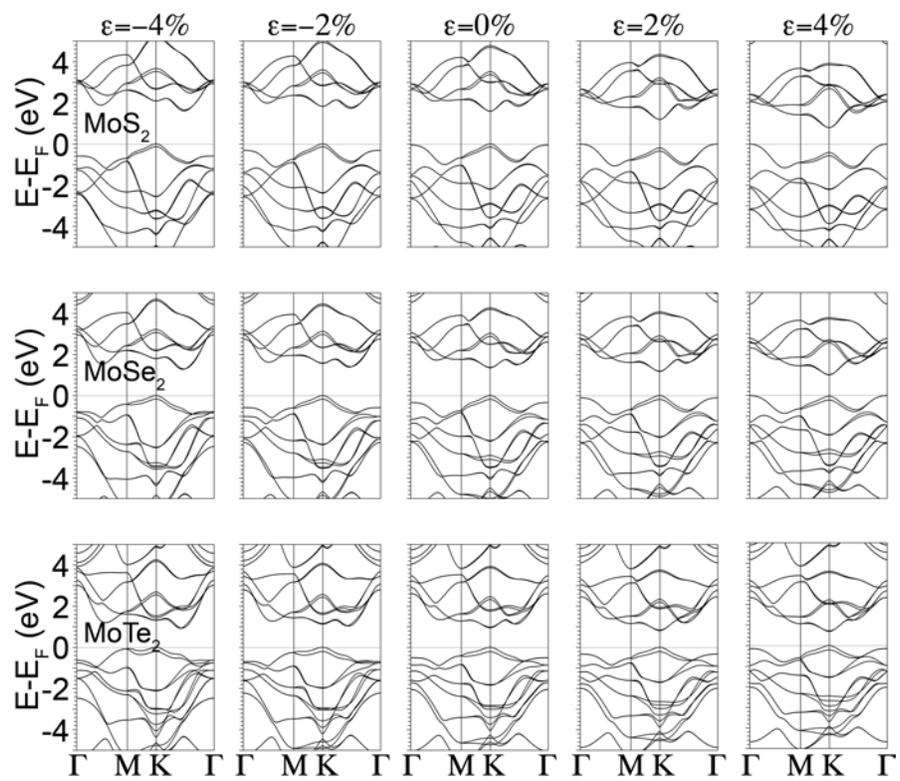

Figure S6. Band structure with SO interaction for transition-metal dichalcogenide single-layer systems with a trigonal prismatic coordination. Systems shown consists of Mo(S,Se,Te)$_2$.

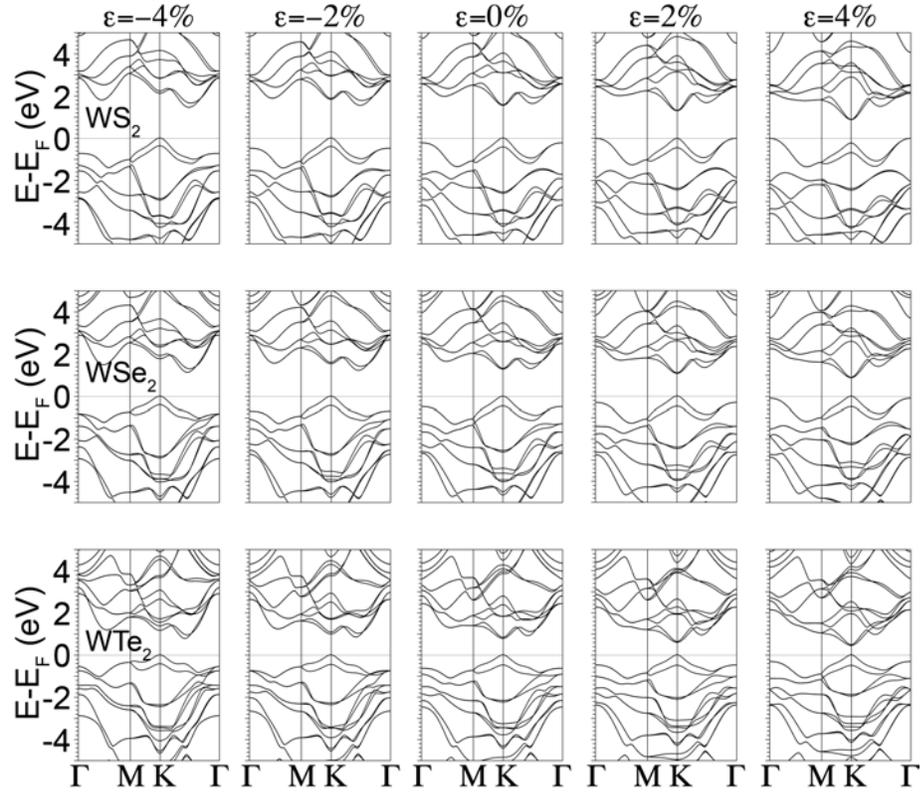

Figure S7. Band structure with SO interaction for transition-metal dichalcogenide single-layer systems with a trigonal prismatic coordination. Systems shown consists of W(S,Se,Te)$_2$.